\documentclass[aps,prl,reprint,nofootinbib,superscriptaddress]{revtex4-1}
\usepackage{blindtext}
\usepackage[english]{babel}
\usepackage{graphicx}
\usepackage{float}
\usepackage{xcolor}

\usepackage{amsmath} 
\usepackage{amssymb}
\usepackage{amsfonts}
\usepackage{hyperref} 
\hypersetup{pdfborder={0 0 0}} 
\usepackage{soul}

\usepackage{pgfplots}
\usepackage{tikz}
\usepackage{psfrag}%

\usepackage{xspace}
\newcommand{\ket}[1]{\ensuremath{|#1\rangle}\xspace}
\newcommand{\bra}[1]{\ensuremath{\langle #1|}\xspace}

\usepackage{mhchem}

\usepackage{tikz}
\usetikzlibrary{matrix}
\usepackage{bm}
\begin{document}
\title{Creating triple-NOON states with ultracold atoms via chaos-assisted tunneling}
\author{G. Vanhaele}
\affiliation{CESAM Research Unit, University of Liege, 4000 Li\`ege, Belgium}
\author{A. B\"acker}
\author{R. Ketzmerick}
\affiliation{Technische Universit\"at Dresden, Institut f\"ur Theoretische Physik and Center for Dynamics, 01062 Dresden, Germany}
\author{P. Schlagheck}
\affiliation{CESAM Research Unit, University of Liege, 4000 Li\`ege, Belgium}
\date{\today}

\begin{abstract}
Triple-NOON states are superpositions of the
form $e^{i \varphi_1} |{N,0,0}\rangle + e^{i \varphi_2} |{0,N,0}\rangle + e^{i \varphi_3} |{0,0,N}\rangle$ involving $N$ bosonic quanta distributed over three modes. We theoretically show
how such highly entangled states can be generated with interacting
ultracold bosonic atoms in a symmetric three-site lattice. The
basic protocol consists in preparing all atoms on one site of the
lattice and then letting the system evolve during a specific time such that collective tunneling of the atoms to the other two
sites takes place. The key point put forward here is that this evolution time can be reduced by several orders of magnitude via the
application of a periodic driving of the lattice, thereby
rendering this protocol feasible in practice. This driving is suitably
tuned such that classical chaos is generated in the entire accessible
phase space except for the Planck cells that host the states
participating at the above triple-NOON superposition. Chaos-assisted
tunneling can then give rise to a dramatic speed-up of this collective
tunneling process, without significantly affecting the purity of this
superposition. A triple-NOON state containing $N = 5$
particles can thereby be realized with $^{87}$Rb atoms on time scales of
the order of a few seconds.
\end{abstract}

\maketitle

Entanglement is a key resource for quantum
information \cite{Nielsen2010} and lies at the heart of various protocols in the context of quantum communication
and quantum computation. Quantum states that feature a high degree of
entanglement are therefore of great interest. A particularly prominent class of highly entangled
states are NOON states \cite{Boto2000,Dowling2008}, $e^{i \varphi_1} \ket{N,0} + e^{i \varphi_2} \ket{0,N}$ (with arbitrary phases $\varphi_1,\varphi_2$), involving $N$
	bosonic quanta that are distributed over two modes. These
	Schr\"odinger-cat states have interesting
	applications in particular for the purpose of quantum metrology \cite{Dowling2008,Simon2017,Pezze2018}.
	They are notoriously difficult to generate if the number $N$ of
	involved quanta is large, even though impressive results were recently
	obtained with the two polarization states of photons \cite{Israel2014}, two optical paths of photons \cite{Afek2010}, the nuclear spin of molecules \cite{Jones2009}, qubits in superconducting circuits \cite{Song2017} and phonons in ion traps \cite{Zhang2018}.
	
	The technical complexity that is inherent in the generation schemes for
	NOON states is certainly a reason why this concept was hardly ever
	considered beyond the paradigm of two-mode systems. A notable exception
	is Ref.~\cite{Yao2021} where the production of \emph{triple-NOON states} of the
	form $e^{i \varphi_1} \ket{N,0,0} + e^{i \varphi_2} \ket{0,N,0} + e^{i \varphi_3} \ket{0,0,N}$ was theoretically investigated for a gas of ultracold
	bosonic atoms. A key ingredient in those bosonic quantum gases is the presence of atom-atom interaction, which induces a spectral separation of NOON
	states from other states with identical total population of the modes.
	This effect was leveraged in a number of theoretical proposals
	for generating two-mode NOON states with bosonic atoms
	\cite{Cirac1998,Sorensen2001,Gordon1999,Micheli2003,Mahmud2003,Zibold2010,Teichmann2007,Carr2010, Grun2022}. It also comes into play in Ref.~\cite{Yao2021} where an adiabatic
	transition to the triple-NOON state was investigated in a
	symmetric three-site system considering bosonic atoms that exhibit an
	attractive mutual interaction. This proposal is intriguing but may be challenging to implement in practice, as extremely low
	temperatures as well as very slow parameter variations would be
	required.
	
	In this paper, we propose a different strategy to generate triple-NOON
	states: within a symmetric three-site system, we consider an initial state where all atoms
	are located on one of the three sites. The NOON state will then
	naturally emerge from collective tunneling of the atoms to the other two
	sites, provided the system is in a self-trapping parameter regime \cite{Smerzi1997,Milburn1997,Leggett2001,Franzosi2001,Franzosi2003} where
	sequential tunneling is inhibited by a sufficiently strong interaction. While the involved collective tunneling times
	are normally prohibitively long to implement such a protocol in
	practice, they can be drastically reduced by means of a suitably tuned
	periodic driving of the three-site system, generalizing a concept that was recently proposed for NOON states in two-site systems \cite{Vanhaele2021} (see also Ref.~\cite{Strzys2008} in this context). Owing to chaos-assisted tunneling \cite{Lin1990, Bohigas1993, BohigasTom1993, Tomsovic1994, Leyvraz1996}, this
	allows one to create triple-NOON states with populations of
	the order of $N=5$ particles, on time scales that are 
	experimentally accessible and feature more favorable scalings with $N$
	than what can be obtained with purely static protocols.

Let us consider a symmetric Bose-Hubbard trimer described by the  Hamiltonian \cite{Nemoto2000,Franzosi2001,Franzosi2003,Hiller2009,Cao2015, Bychek2020},
\begin{equation}\label{H_BH3_d_ladder}
\hat H(t) = -\tilde J(t) \sum_{l=0}^{2} (\hat a_l^\dagger \hat a_{l+1} + \hat a_{l+1}^\dagger \hat a_l )
+ \dfrac{U}{2} \sum_{l=0}^{2}  \hat a_l^\dagger \hat a_l^\dagger \hat a_l \hat a_l,
\end{equation}
with $\hat a_3 \equiv \hat a_0$. This trimer can represent a model for $N$ ultracold atoms that are contained within an isolated triangle of an optical Kagom\'e lattice \cite{Santos2004,Damski2005} featuring one orbital per lattice site, where $U$ is the on-site atom-atom interaction strength. It is subjected to a periodic modulation of its inter-site hopping parameter \cite{Pieplow2019}
\begin{equation}
\tilde J(t)=J+\delta \cos(\omega t),
\end{equation}
 with $J$ the hopping in absence of the driving, $\delta$ the amplitude, and $\omega$ the frequency of the driving (see, e.g., Refs.~\cite{Lignier2007,Sias2008,Kierig2008,Zenesini2009,Zenesini2009_2} for pioneering experimental realizations of driven lattices and Floquet engineering).

\begin{figure}[!h]
	\centering
%
	\includegraphics[width=0.45\textwidth]{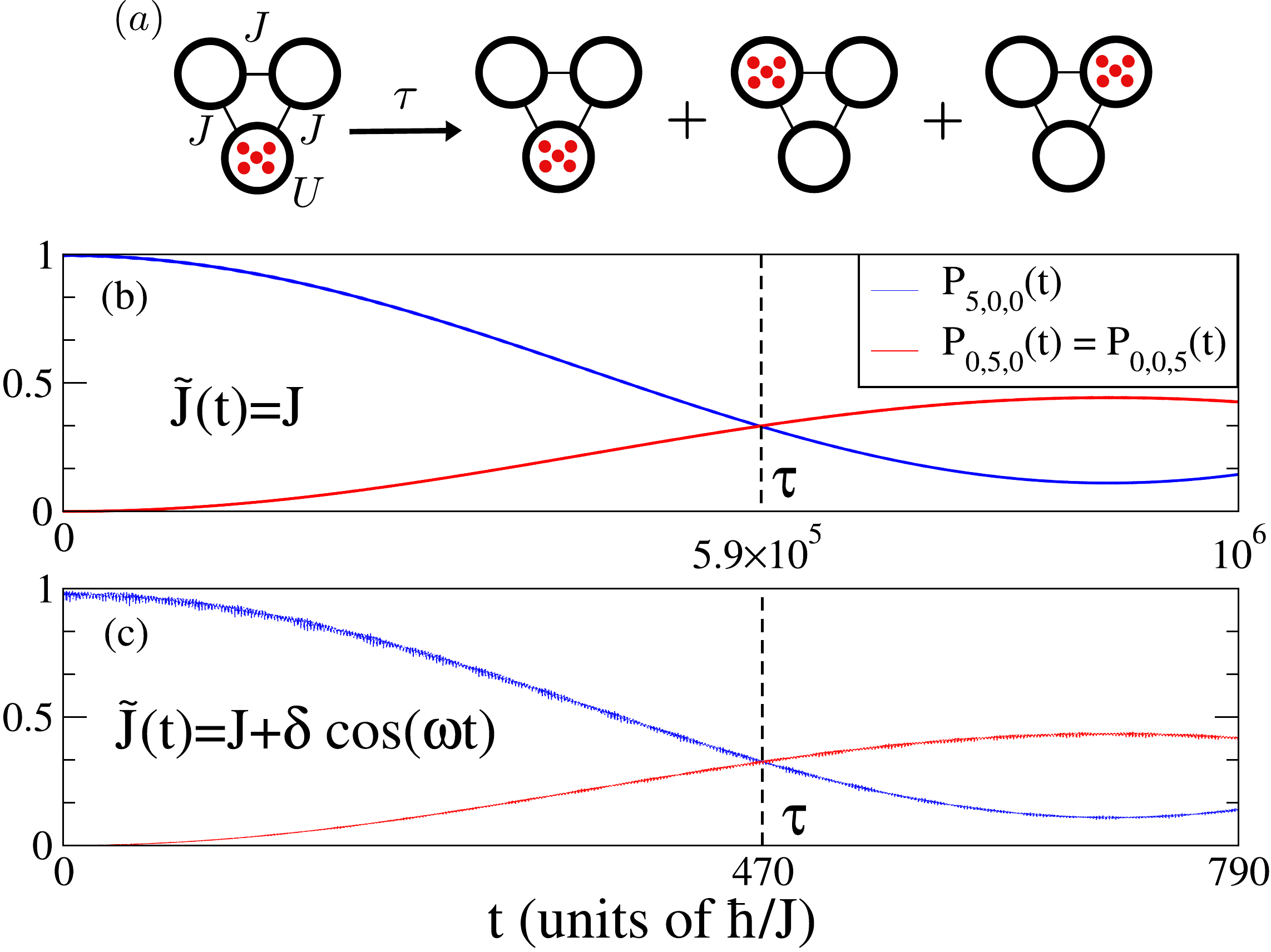}
	\caption{(a) Schematic representation of the Bose-Hubbard trimer. Free time evolution, starting from the initial state $\ket{N,0,0}$, gives rise to the triple-NOON state after the evolution time $\tau$. (b) Time-dependent detection probabilities \eqref{proba_def} of the initial state $\ket{N,0,0}$ [blue (dark gray) line] and its symmetric counterparts $\ket{0,N,0}$, $\ket{0,0,N}$ [red (light gray) line], computed in the self-trapping regime for $N=5$ and $U/J=20$. The triple-NOON time, when the perfectly balanced entangled state between $\ket{N,0,0}$, $\ket{0,N,0}$ and $\ket{0,0,N}$ is reached, is evaluated as $\tau=5.9 \times 10^5\hbar/J$. (c) Same as (b) but in the presence of a time-periodic modulation of the hopping with the amplitude $\delta=J$ and the frequency $\omega=60 J/\hbar$. Note the drastic reduction of the triple-NOON time, $\tau =470\hbar/J$, as compared to the undriven case.}
	\label{Fig_BH3_tripleNOON_proba}
\end{figure}

Let us first assume that there is no driving, i.e., $\tilde J(t)=J$. For large $NU/J$, with $N$ the number of atoms, transfers of particles between the sites of the trimer are in principle possible but can be strongly suppressed due to the mismatch between the chemical potentials on the sites. This is specifically the case for an initial state $\ket{N,0,0}$ which will be subjected to self-trapping \cite{Smerzi1997,Milburn1997,Leggett2001,Franzosi2001,Franzosi2003}. On the other hand, due to the symmetry between the states $\ket{N,0,0}$, $\ket{0,N,0}$ and $\ket{0,0,N}$, a population transfer between the sites [see Fig.~\ref{Fig_BH3_tripleNOON_proba}(a)] will nevertheless take place, albeit on a very long time scale.

More quantitatively, the three-level dynamics can be modeled by a matrix, 
\begin{equation}
H_{\rm eff} = 
\begin{pmatrix}
E & V & V \\ 
V & E & V \\
V & V & E 	
\end{pmatrix},
\end{equation}
that represents the decomposition of the Hamiltonian (\ref{H_BH3_d_ladder}) in the basis $\{\ket{N,0,0}, \ket{0,N,0},\ket{0,0,N}\}$.
The unperturbed energies are given by $E=\frac U 2 N(N-1)$ and $V$ is the effective coupling matrix element between the quasimodes. The eigenvalues are $E-V$, $E-V$, and $E+2V$ with the splitting $\Delta \epsilon=3V$. The time evolution of the system, initially prepared in $\ket{N,0,0}$, gives rise to Rabi-like oscillations of the form
\begin{align}
\ket{\phi(t)} =&
\dfrac 1 3 \left(2 e^ {i\Delta\epsilon  t/(2\hbar)} + e^ {-i\Delta \epsilon  t/(2\hbar)}  \right) \ket{N,0,0} \nonumber\\
&-\dfrac{2i}{3} \sin\left(\dfrac{\Delta\epsilon }{2\hbar} t\right) \left(\ket{0,N,0}+\ket{0,0,N} \right).
\label{phi_t}
\end{align}
The detection probabilities,
\begin{equation}\label{proba_def}
P_{n_0,n_1,n_2}(t)=|\bra{n_0,n_1,n_2} \phi(t)\rangle|^2,
\end{equation}
(to be yielded, in practice, with quantum gas microscopes, see, e.g., Ref.~\cite{Kaufman16})
are the same for $\ket{0,N,0}$ and $\ket{0,0,N}$. At the triple-NOON time,
\begin{equation}\label{tNOON_time}
\tau=\dfrac{2\pi\hbar}{3|\Delta \epsilon|},
\end{equation}
 one has a perfectly balanced superposition between the three quasimodes at stake, i.e., $P_{N,0,0} (\tau)=P_{0,N,0} (\tau)=P_{0,0,N} (\tau) =  1/3$.

Figure \ref{Fig_BH3_tripleNOON_proba}(b) shows the time evolution of these detection probabilities for $N=5$ and $U/J=20$, computed via numerical diagonalization of the Bose-Hubbard Hamiltonian \eqref{H_BH3_d_ladder}. The triple-NOON time is evaluated as $\tau=5.9\times 10^5 \hbar/J$. To obtain an idea of what this means in practice, we consider the case of $\ce{^87Rb}$ atoms, characterized by a mass $m=1.443\times 10^{-25}$ kg and an s-wave scattering length $a_s=5.313$ nm, within an optical lattice that is produced by lasers of wavelength $\lambda=1064$ nm. This yields $\hbar/J= 4.4\times 10^{-3}$ (see Ref.~\cite{Vanhaele2021} for more details), from which we infer $\tau=2600$ s. This is prohibitively long in comparison to the typical lifetime $\sim 10$ s of a condensate in an optical lattice \cite{Andersen2008}.

This very large value of the triple-NOON time is a consequence of the fact that the transition between the states $\ket{N,0,0}$, $\ket{0,N,0}$, and $\ket{0,0,N}$ arises due to collective tunneling process involving the simultaneous participation of all particles. It is, however, known for a few decades that such tunneling processes can be strongly enhanced via the application of an external periodic driving to the system at hand, to the extent that this time characterizing this process can decrease by several orders of magnitude \cite{Lin1990, Bohigas1993,BohigasTom1993, Tomsovic1994,Leyvraz1996, Brodier2001,Brodier2002,Eltschka2005,Keshavamurthy2005JCP,Schlagheck2006, Mouchet2006,Backer2009,Lock2010, Schlagheck2011,Kullig2016, Mertig2016, Fritzsch2017,  Fritzsch2019}. Note that this driving does not need to be so strong as to appreciably alter the characteristics of the states $\ket{N,0,0}$, $\ket{0,N,0}$, and $\ket{0,0,N}$ between which tunneling takes place; its primary function is to induce perturbative coupling matrix elements to other states (such as $\ket{N-1,1,0}$) that are more strongly coupled to a  number of other states of the system by the presence of the driving, corresponding to chaos-assisted tunneling \cite{Lin1990, Bohigas1993, BohigasTom1993, Tomsovic1994, Leyvraz1996} (or exhibit stronger intrinsic tunneling rates, which would be resonance-assisted tunneling \cite{Brodier2001, Brodier2002})

This phenomenon can be leveraged to speed up the creation of triple-NOON states. Figure \ref{Fig_BH3_tripleNOON_proba}(c) shows a numerical simulation of the time evolution of the detection probabilities for $\delta=J$ and $\omega=60\hbar/J$, which was obtained through a diagonalization of the corresponding Floquet Hamiltonian \cite{Shirley1965, Sambe1973}. We find here a drastically reduced triple-NOON time amounting to $\tau=470\hbar/J$. For a typical optical lattice filled with $\ce{^87Rb}$ (see above), the corresponding triple-NOON time in laboratory units is $\tau=2.1$ s. This reduction of three orders of magnitude can enable an experimental observation of the triple-NOON state.

The chaos-assisted tunneling phenomenon that is at work here can be qualitatively and quantitatively understood through an analysis of the corresponding classical dynamics. The latter is given in terms of the discrete Gross-Pitaevskii equation,
\begin{equation}\label{GP_BH3_d_psi}
i \hbar \dfrac{d \psi_l}{dt} = - \tilde J(t)(\psi_{l-1}+\psi_{l+1}) +  U |\psi_l|^2 \psi_l,
\end{equation}
for $l=0,1,2$ (with $\psi_{-1} \equiv \psi_2$ and $\psi_3 \equiv \psi_0$).
 Here a condensate amplitude $\psi_l=\sqrt{n_l+1/2}\ e^{i\theta_l}$ is associated with each site where $n_l$ and $\theta_l$ are respectively the number of particles and the phase on site $l$.
This Gross-Pitaevskii equation \eqref{GP_BH3_d_psi} is generated by the classical Hamiltonian (see also Ref.~\cite{Mossmann2006})
\begin{align}
H(\phi_0, \phi_1,& z_0, z_1, t)  = U z_1^2 + \dfrac{3U}{4} \left(z_0- {\tilde N}/ 3 \right)^2  \nonumber \\
 -\tilde J(t)& \left( \sqrt{2 z_0 (\tilde N-z_0+2z_1)} \ \cos\left(\phi_0-  \phi_1/2\right) \right. \nonumber \\
& + 2 \sqrt{(\tilde N-z_0)^2/4- z_1^2} \ \cos(\phi_1)  \label{V_phi_z_t}\\ 
&\left. + \sqrt{2 z_0 (\tilde N-z_0-2z_1)} \ \cos\left(\phi_0+  \phi_1/2\right) \right),  \nonumber
\end{align}
with $\tilde N=N+3/2$. This time-periodic Hamiltonian defines a stroboscopic map in a phase space of four dimensions (using total particle number conservation) with coordinates
\begin{align}
\begin{aligned}
&\phi_0= \theta_0 - (\theta_1+\theta_2)/2, \\
&\phi_1= \theta_1 -\theta_2,
\end{aligned}
\qquad \qquad 
\begin{aligned}
& z_0= n_0 + 1 /2,\\
&z_1 = (n_1-n_2)/2 ,
\end{aligned}
\end{align}
where $z_0$ represents the population of site 0, and $z_1$ is the population imbalance between site 1 and 2.

\begin{figure}[!h]
	\includegraphics[width=\linewidth]{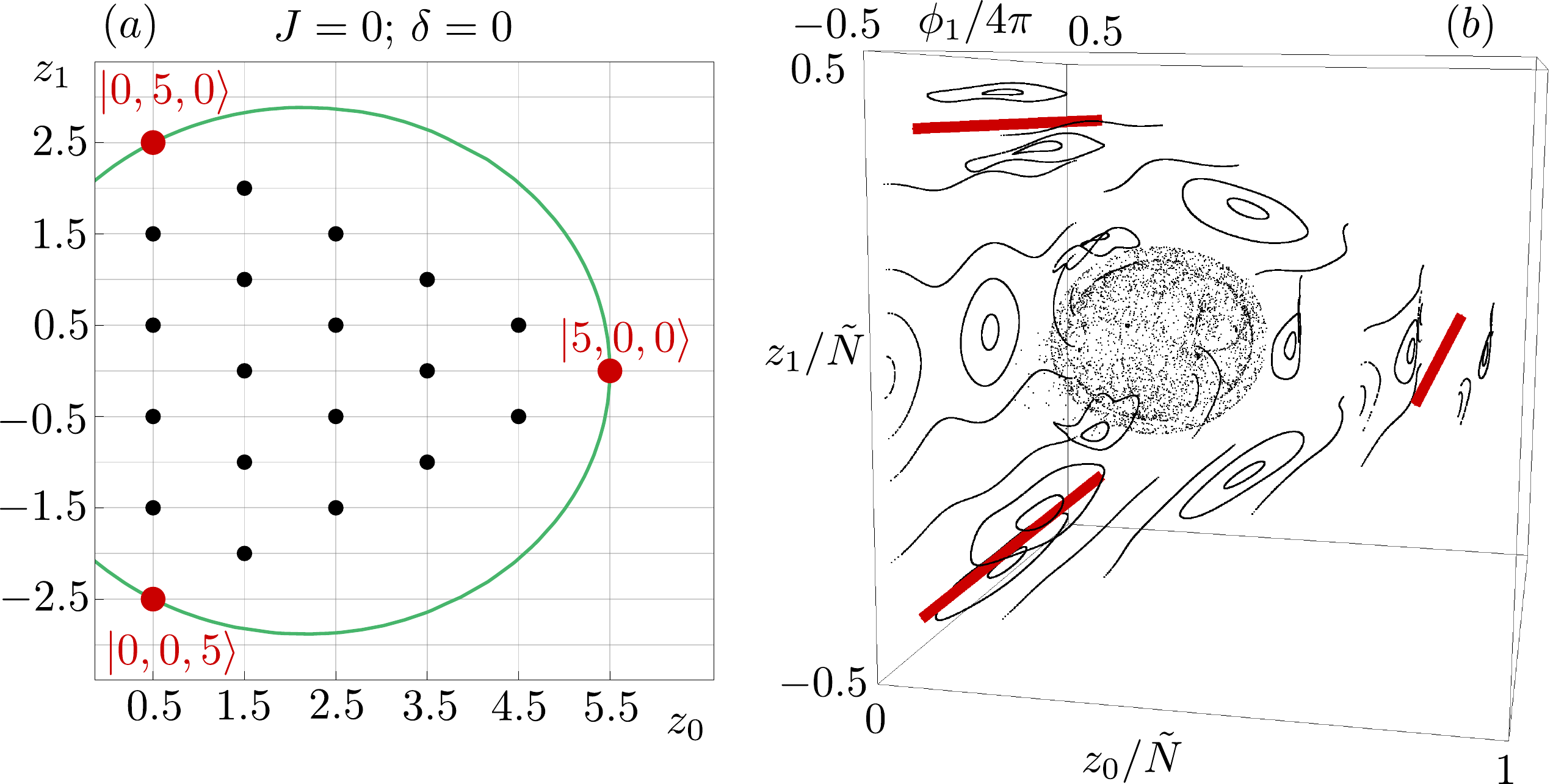}

	\caption{(a) Location of the quasimodes $\ket{n_0,n_1,n_2}$ with $\sum_i n_i =5$ (dots) in action space. The green ellipse indicates the unperturbed energy shell, associated with the disconnected Bose-Hubbard trimer, that contains the quasimodes $\ket{5,0,0}$, $\ket{0,5,0}$, and $\ket{0,0,5}$. (b) 3D phase-space slice of the connected ($J>0$) but undriven ($\delta=0$) Bose-Hubbard trimer obtained for $N=5$ and $U/J=20$. The straight red lines mark the location of the quasimodes $\ket{5,0,0}$, $\ket{0,5,0}$, and $\ket{0,0,5}$, in close analogy with panel (a). The phase-space structure is mainly regular, with some chaotic regions in the center.	
	}
	\label{Fig_grid}
\end{figure}

The four-dimensional phase space can be visualized using a three-dimensional (3D) phase-space slice \cite{Richter2014,Lange2014}, where the coordinates $(\phi_1,z_0, z_1)$ of the stroboscopic time series (evaluated for $\omega t=0$ mod $2\pi$) are displayed whenever the slice condition $|\phi_0|\leqslant \varepsilon$ for some $\varepsilon>0$ is satisfied.
If the system is time independent, the slice can be performed with $\varepsilon=0$ to produce a Poincar\'e section of the continuous time dynamics.  
Figure \ref{Fig_grid}(b) shows such a 3D phase-space slice for the undriven three-site Bose-Hubbard trimer at $U/J=20$.
The phase space is characterized by three pendulum-like structures, corresponding to the three cosine terms in Eq.~\eqref{V_phi_z_t}, as it is visible, e.g., in the plane $z_0=0$.
 At their intersection, there exists a resonance junction \cite{Hal1999,EftHar2013,Firmbach2019} centered about $(z_0,z_1)= ( {\tilde N}/ 3, 0)$,  giving rise to a chaotic bubble.

Straightforward semiclassical torus quantization rules yield the location of the quasimodes in the action space spanned by $z_0$ and $z_1$, as shown in Fig.~\ref{Fig_grid}(a). Those quasimodes correspond to the Fock states $|n_0, n_1, n_2\rangle$, which are the eigenstates of the disconnected Bose-Hubbard trimer (i.e., for zero hopping $\tilde{J} \equiv 0$). They acquire nonvanishing tunnel couplings in the presence of inter-site hopping. The quasimodes $|N,0,0\rangle$, $|0,N,0\rangle$, $|0,0,N\rangle$ that participate at the NOON superposition are situated near the corners in the action space and are energetically isolated from other quasimodes with identical particle number $N$. Their mutual coupling therefore involves barrier tunneling taking place on very long time scales.

\begin{figure}[!h]
	\centering
	\includegraphics[width=0.5\linewidth]{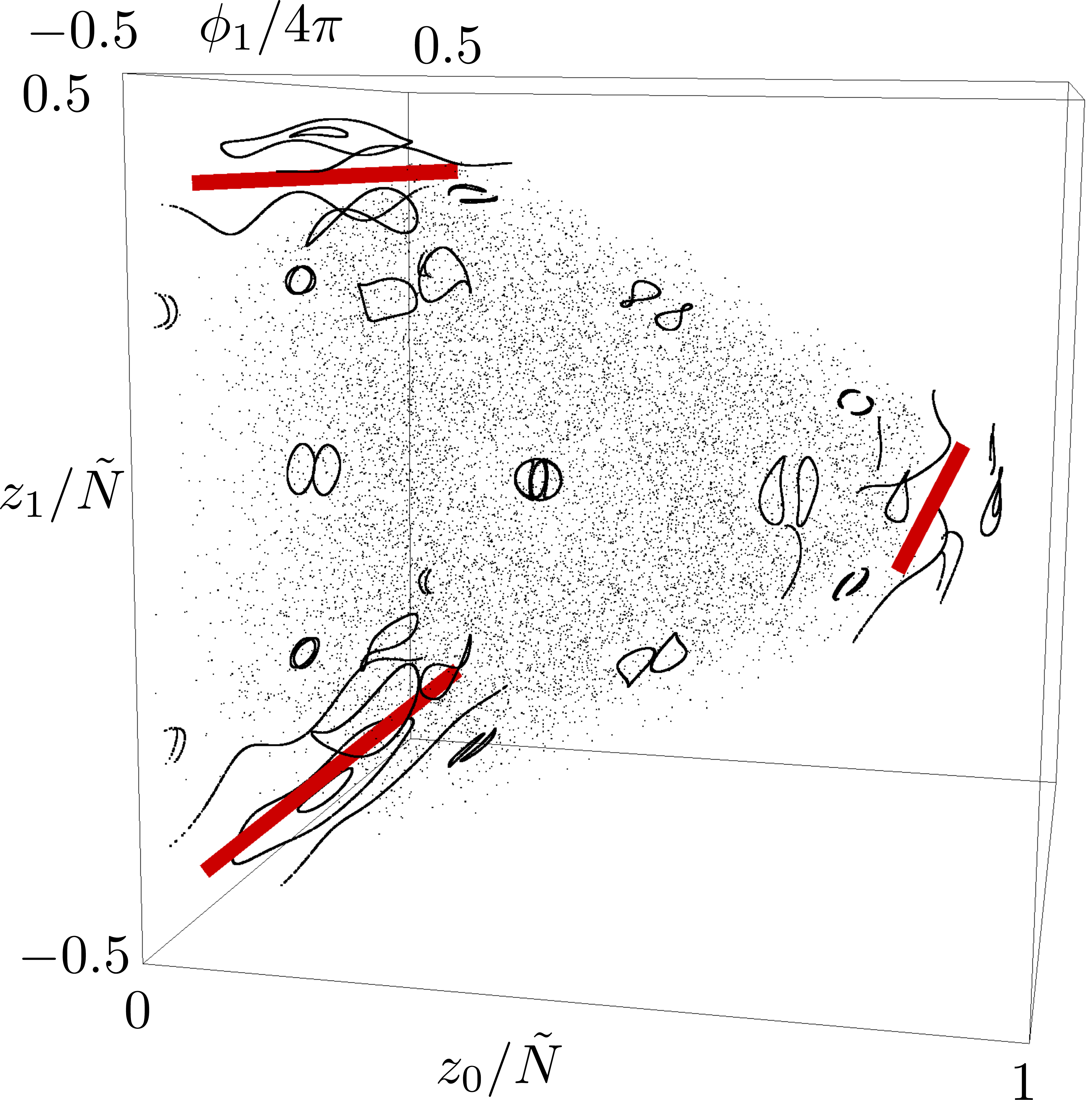}
	\caption{
		Same as Fig.~\ref{Fig_grid}(b) but for the case of a periodically-driven Bose-Hubbard trimer, with the driving parameters $\delta=J$ and $\hbar\omega=60J$. The phase-space structure is dominantly chaotic, exhibiting a few islands of regular motion. Predominant regular dynamics is still encountered in these corners of the phase space that host the quasimodes $\ket{5,0,0}$, $\ket{0,5,0}$, and $\ket{0,0,5}$ involved in the triple-NOON superposition.
	}
	\label{Fig_PS_BH3chaos}
\end{figure}

The presence of a periodic driving effectively lifts this energetic isolation of the NOON quasimodes and induces various additional couplings between the quasimodes of the system. Correspondingly, the classical dynamics becomes more chaotic. This is illustrated in Fig.~\ref{Fig_PS_BH3chaos} which shows a 3D phase-space slice (with $\varepsilon  = 10^{-3}$) for the parameters $U/J=20$, $\delta=J$, and $\hbar\omega = 60 J$. This latter value for the frequency was deliberately chosen in order to approximately match the characteristic oscillation frequencies of the undriven Bose-Hubbard trimer near the central region of the phase space. A still rather moderate value for the driving amplitude is then sufficient to turn this central phase-space region into a large chaotic sea, interspersed by tiny islands of regular motion. At the same time, again thanks to the deliberate choice of the driving frequency, the three corners that host the quasimodes participating at the triple-NOON superposition are only marginally affected by this driving.

A semiclassical theory of dynamical tunneling in such high-dimensional mixed regular-chaotic systems, which would allow one to quantitatively predict the corresponding tunneling rates based on purely classical information, is still object of current fundamental research investigations \cite{Karmakar2018, Firmbach2019, Karmakar2020IVR, Karmakar2020unimolecular, Karmakar2021}. From a qualitative point of view, the experience acquired from the studies of similar tunneling problems in mixed regular-chaotic systems with two effective degrees of freedom (Ref.~\cite{Vanhaele2021} in particular) tells us that this particular configuration of the phase space is about optimal for maximizing tunneling between the quasimodes $|N,0,0\rangle$, $|0,N,0\rangle$, $|0,0,N\rangle$ without significantly admixing those modes to other states of the system. Indeed, judging from Fig.~\ref{Fig_PS_BH3chaos} in comparison with the action space shown in Fig.~\ref{Fig_grid}(a), the extent of each regular corner hosting one of those quasimodes roughly corresponds to the size of a Planck cell, which indicates that all other quasimodes are more or less strongly affected by the chaotic part of the phase space. A straightforward application of random matrix theory for this chaotic manifold \cite{Tomsovic1994,Leyvraz1996} yields $|v|^2/(\hbar^2\omega)$ as characteristic scale for the tunneling rate where $v$ is the largest driving-induced matrix element between a regular state (i.e.,~$|N,0,0\rangle$, $|0,N,0\rangle$, $|0,0,N\rangle$) and any of the other quasimodes (such as $|N-1,1,0\rangle$) \cite{Eltschka2005}.

\begin{figure}[!h]
	\centering
	\includegraphics[width=0.48\textwidth]{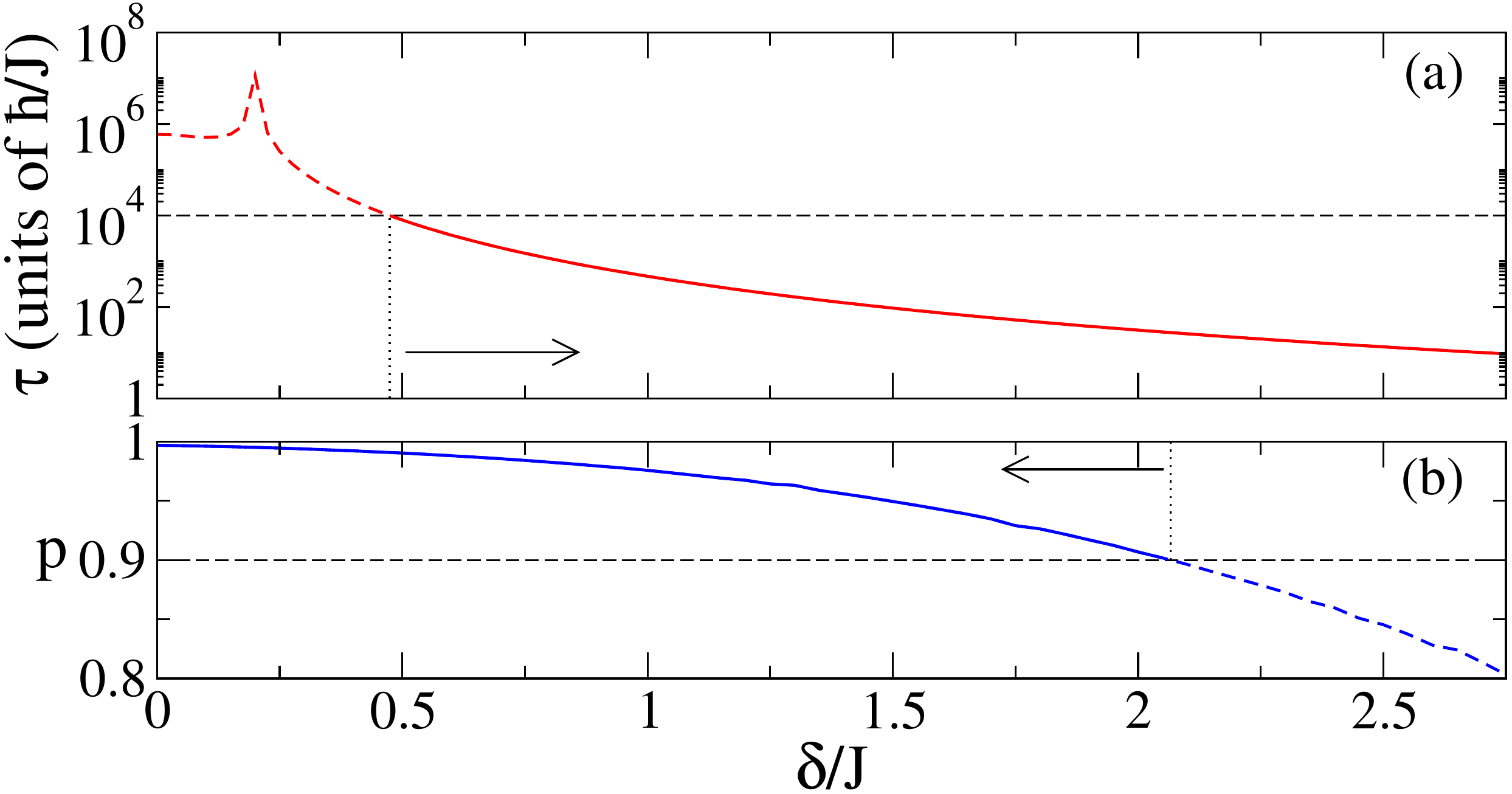}
	\caption{The triple-NOON time (a), defined in Eq.~\eqref{tNOON_time}, and the purity (b), defined in Eq.~\eqref{purity_Eq}, as a function of the amplitude $\delta$ of the driving for $N=5$, $U/J=20$ and the driving frequency $\omega =60 J/\hbar$. While the tunneling process towards the triple-NOON state becomes faster with increasing driving amplitude, the purity decreases. The definitions of tolerance thresholds for the triple-NOON time, which must, e.g., not exceed $\tau_{\rm max}=10^4 \hbar/J$ [dashed line in (a)], and the purity, which must, e.g., be at least $p_{\rm min}=0.9$ [dashed line in (b)], allows one to identify an acceptable parameter range for the driving amplitudes, as indicated by the arrows.
	}
	\label{Fig_tau_d}
\end{figure}

Fixing the driving frequency in the above manner, such that the driving resonantly couples states located in the central part of the phase space without affecting the corners that host the NOON superposition, allows one to efficiently reduce the time scale needed for the production of the NOON state using a rather moderate driving amplitude. This is shown in Fig.~\ref{Fig_tau_d}(a) which displays the numerically computed triple-NOON time as a function of the driving amplitude. If we introduce, somewhat arbitrarily, $\tau_{\rm max}=10^4 \hbar/J$ (corresponding to roughly 44 s in the case of $\ce{^87Rb}$ atoms in optical lattices, see above) as maximally acceptable threshold value for $\tau$, then the driving amplitude has to be chosen larger than $0.5J$.

The triple-NOON time continues to decrease with further increasing driving amplitude, but the quality of the NOON superposition starts to significantly degrade for $\delta>J$, due to increasingly important admixtures of components $\ket{n_0,n_1,n_2}$ that are unrelated to the NOON state. To quantify this, we introduce the purity,
\begin{equation}\label{purity_Eq}
p= \dfrac 1 T \int_0^T dt \left[ P_{N,0,0}(t)+  P_{0,N,0}(t)+  P_{0,0,N}(t)\right],
\end{equation}
where the temporal average is performed over a sufficiently long time (e.g.~$T=100\times 2\pi/\omega$) in order to level out oscillatory effects due to micromotions. As shown in Fig.~\ref{Fig_tau_d}(b), the purity is very close to unity in the undriven Bose-Hubbard trimer and decreases monotonously with increasing driving amplitude. If we define, again somewhat arbitrarily, $p_{\rm min}=0.9$ as minimally acceptable threshold value for $p$, then we have to choose $\delta<2.1 J$. Quite naturally, $\delta=J$ is therefore a near-optimal choice for the driving amplitude in order to produce triple-NOON states with high purity ($p=0.976$) on reasonable time scales ($\tau=470 \hbar/J$, corresponding to $\tau=2.1$ s for $\ce{^87Rb}$, see above).
Figure \ref{Fig_tau_d} also demonstrates robustness with respect to small parameter variations, which facilitates the implementation of this driving protocol.

The above consideration can be generalized for different total particle
numbers $N$ while keeping the same classical phase-space structure. To
this end, the interaction strength $U$ has to be rescaled such that
$\tilde{N} U = (N + 3/2)U$ is fixed while keeping $J$ and $\omega$ constant.
As $\tilde{N}$ effectively represents the inverse Planck constant, an
exponential increase of the triple-NOON time with $\tilde{N}$ is to be
expected in the undriven trimer. This exponential increase can be very
favorably amended by the presence of the driving. Since the size
of a Planck cell shrinks with increasing $N$, slightly larger values for
the driving amplitude can be employed for $N$ larger than 5, in order to
achieve maximal chaos conditions outside the Planck cells hosting
$|N,0,0\rangle$, $|0,N,0\rangle$, $|0,0,N\rangle$ while keeping those three quasimodes
essentially unaffected. As shown in Table \ref{table_Np}, the thereby obtained
near-optimal values for the triple-NOON time still increase with $N$ but
much less dramatically than in the absence of the driving.
\begin{table}[!h]
	\centering
	$\begin{array}{c|c|c|c}
	           & N=5 & N=7 & N=9 \\
	           \hline
	 \delta =0 &  5.9 \times 10^5   & 1.0 \times 10^9    &  1.9 \times 10^{12}  \\
	 \delta \neq 0 & 4.7 \times 10^2 &  4.8 \times 10^3  & 4.3 \times 10^6
	\end{array}$
	\caption{Triple-NOON times $\tau$ (in units of $\hbar/J$) for different particle numbers, keeping fixed $\tilde N U/J=130$,
		in the undriven (top row) and the periodically-driven case (bottom row) with  $\hbar\omega=60J$.
		Optimal driving amplitudes are respectively found at $\delta=J$, $1.3J$, $1.5J$ for $N=5$, $7$, $9$.}
	\label{table_Np}
\end{table}

In conclusion, we show that a triple-NOON state can be realized with ultracold bosonic atoms in a symmetric three-site lattice that is exposed to a suitably tuned periodic driving. Chaos-assisted tunneling is the key semiclassical mechanism that is leveraged in this context, giving rise to a drastic reduction of the collective tunneling time that one has to let the system evolve in order to obtain the equiprobable superposition between $|N, 0, 0\rangle$, $|0, N, 0\rangle$, and $|0, 0, N \rangle$.
The analysis of underlying classical dynamics enables one to determine near-optimal driving parameters for a rather rapid generation of triple-NOON states with a very high purity. Those parameters can form the starting point for the application of quantum control techniques to further optimize the driving protocol, which may allow one to further boost the involved collective tunneling process. 

While tunneling times smaller than the typical life time of the atomic gas can thereby be reached, other important technical challenges are still to be faced in order to create triple-NOON states in practice. Specifically, a decoherence-free environment has to be maintained throughout the collective tunneling process. Heating due to the driving-induced population of excited Wannier modes has to be avoided, possibly through fine tuning of the driving frequency such that the driving resonantly couples the ground mode into a band gap of the three-well lattice. And very stringent symmetry requirements, both in the configuration of the lattice and in the driving, have to be implemented and maintained (which can possibly be achieved by replacing the static lattice under external driving with a three-site time crystal \cite{Sacha2015} featuring intrinsic symmetry). Further studies are certainly required to investigate those additional challenges in more detail, and to extend this study to other types of atomic gases, e.g.\ involving dipolar interactions \cite{Lahaye2010,Wilsmann2018}.
 
\begin{acknowledgments}
We thank Felix Fritzsch, Markus Firmbach, and Srihari Keshavamurthy for inspiring discussions. Financial support from the Belgian F.R.S.-FNRS (FNRS aspirant grant for G.V.) is gratefully acknowledged. Funded by the Deutsche Forschungsgemeinschaft (DFG, German Research Foundation) -- 290128388; 497038782.
\end{acknowledgments}

\bibliographystyle{apsrev4-1} 

\bibliography{biblio} 

\end{document}